\documentclass{article}
\usepackage{arxiv}
\usepackage[utf8]{inputenc}
\usepackage[T1]{fontenc}
\usepackage{amsmath,amssymb,amsthm}
\usepackage{graphicx}
\graphicspath{{figures/}}
\usepackage{float}
\usepackage{wrapfig}
\usepackage{enumitem}
\usepackage{booktabs}
\usepackage{microtype}
\usepackage[numbers,square,sort&compress]{natbib}
\usepackage{tikz}
\usetikzlibrary{decorations.pathreplacing,decorations.markings}
\newif\ifFigureColour
\FigureColourtrue
\ifdefined\FigureMonochrome
    \FigureColourfalse
\fi

\definecolor{figurePaletteBlue}{HTML}{2391FF}
\definecolor{figurePaletteYellow}{HTML}{FFF83F}
\definecolor{figurePaletteAmber}{HTML}{FFCD54}
\definecolor{figurePaletteRose}{HTML}{F70063}

\ifFigureColour
    \colorlet{figureBlue}{figurePaletteBlue!85!black}
    \colorlet{figureBlueFill}{figurePaletteBlue!18}
    \colorlet{figureBlueStrongFill}{figurePaletteBlue!35}
    \colorlet{figureRose}{figurePaletteRose!85!black}
    \colorlet{figureRoseFill}{figurePaletteRose!15}
    \colorlet{figureRoseText}{black}
    \colorlet{figureFocus}{figurePaletteAmber}
    \colorlet{figureFocusBorder}{figurePaletteAmber!80!black}
    \colorlet{figureFocusFill}{figurePaletteAmber!45}
    \colorlet{figureDomainFill}{figureBlueFill}
\else
    \colorlet{figureBlue}{black!70}
    \colorlet{figureBlueFill}{black!7}
    \colorlet{figureBlueStrongFill}{black!18}
    \colorlet{figureRose}{black}
    \colorlet{figureRoseFill}{black!65}
    \colorlet{figureRoseText}{white}
    \colorlet{figureFocus}{black!40}
    \colorlet{figureFocusBorder}{black!40}
    \colorlet{figureFocusFill}{black!12}
    \colorlet{figureDomainFill}{figureBlueFill}
\fi

\tikzset{figure focus border/.style={draw=figureFocusBorder,line width=.7pt}}

\usepackage{todonotes}

\newtheorem{theorem}{Theorem}[section]
\newtheorem{lemma}[theorem]{Lemma}

\newtheorem{corollary}[theorem]{Corollary}
\theoremstyle{definition}

\theoremstyle{remark}

\usepackage[hidelinks]{hyperref}

\newcommand{\enc}{\mathrm{enc}}

\newcommand{\nats}{\mathbb{N}}
\newcommand{\rats}{\mathbb{Q}}
\newcommand{\reals}{\mathbb{R}}

\newcommand{\push}{\mathrm{push}}
\newcommand{\pop}{\mathrm{pop}}

\newcommand{\bs}[1]{\boldsymbol{#1}}

\newcommand{\NEXP}{\textbf{NEXP}}

\newcommand{\dbsat}{\textsc{DegBSat}}

\newcommand{\relu}{\mathrm{relu}}
\newcommand{\trelu}{\mathrm{trelu}}

\title{Degree-bounded Satisfiability of Local MPNNs with Sum Aggregation is
    NEXPTIME-hard}
\author{%
  Marco Sälzer \\
  RPTU University of Kaiserslautern-Landau \\
  Kaiserslautern, Germany \\
  \texttt{marco.saelzer@rptu.de}
}
\date{}
\hypersetup{
  pdftitle={Degree-bounded Satisfiability of Local MPNNs with Sum Aggregation is
    NEXPTIME-hard},
  pdfauthor={Marco Sälzer}
}

\begin{document}
\maketitle

\begin{abstract}
  We prove that the following problem is \textbf{NEXPTIME}-hard: given a message passing
neural network (MPNN) $M$ with local sum aggregation and truncated ReLU activations,
and a degree bound $d$ encoded in unary, decide whether there is a pointed graph
$(G,v)$ with rational node labels and degree at most $d$ that is accepted by $M$.
Hardness already holds for the fixed degree bound $d=3$.
% Additionally, we provide a full Lean formalisation of
% the proof.

\end{abstract}

\section{Introduction}
\label{sec:introduction}

\emph{Message Passing Neural Networks (MPNNs)} are a prominent class of graph neural
networks, used for learning tasks on graphs such as node classification.
Their central idea is simple: in each round, or layer, nodes pass messages based on their states
to their direct neighbours or globally to all nodes. Each node aggregates the incoming messages and
combines the result with its own state to obtain an updated state,
using a learnable feedforward neural network (FNN).

% Resume full-width text immediately below the figure and its caption.
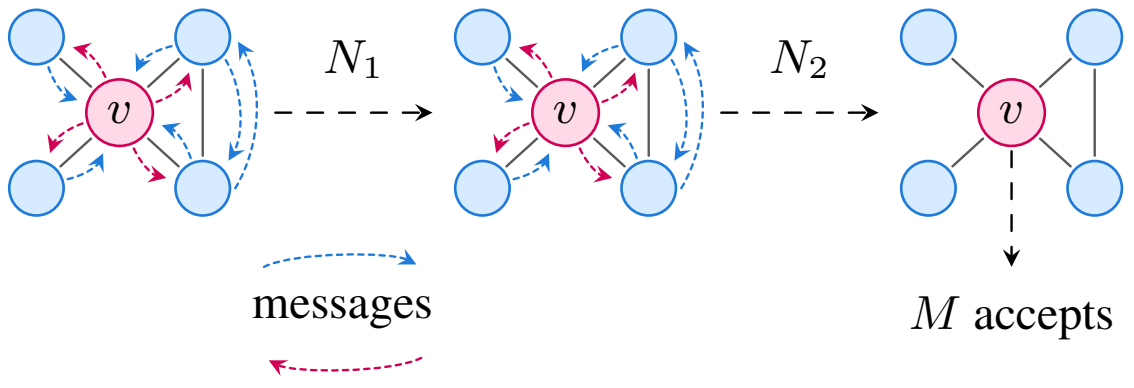
\begin{wrapfigure}[12]{r}{.46\textwidth}
    \vspace{-\intextsep}
    \centering
    \begingroup
% Layout parameters (coordinates in cm); colours remain fixed input labels.
\newcommand{\MPNNRoundSeparation}{2.95}
\newcommand{\MPNNHalfWidth}{.55}
\newcommand{\MPNNHalfHeight}{.50}
\newcommand{\MPNNNodeSize}{.36cm}
\newcommand{\MPNNFocusSize}{.44cm}
\newcommand{\MPNNOutputHeight}{-1.02}
\newcommand{\MPNNMessageBend}{26}
\newcommand{\MPNNNeighbourInnerOffset}{.12}
\newcommand{\MPNNNeighbourOuterOffset}{.18}
\newcommand{\MPNNUpdateStartGap}{.48}
\newcommand{\MPNNLegendHalfWidth}{.52cm}
\newcommand{\MPNNLegendCurveHeight}{.10}

% One copy of the graph. The argument is the state index (0, 1, or 2).
% All copies share the same undirected edges.
\newcommand{\MPNNSnapshot}[1]{%
    \begin{scope}[shift={({#1*\MPNNRoundSeparation},0)}]
        \node[mpnn focus] (mpnn-#1-v) at (0,0) {$v$};
        \foreach \name/\sx/\sy in {a/-1/1,b/1/1,c/-1/-1,d/1/-1} {
            \node[mpnn neighbour] (mpnn-#1-\name)
                at ({\sx*\MPNNHalfWidth},{\sy*\MPNNHalfHeight}) {};
            \draw[mpnn edge] (mpnn-#1-\name) -- (mpnn-#1-v);
        }
        % Connect the upper and lower neighbours on the right.
        \draw[mpnn edge] (mpnn-#1-b) -- (mpnn-#1-d);
        % Show the same message flow in both update rounds.
        \ifnum#1<2
            \foreach \name in {a,b,c,d} {
                \draw[mpnn blue message] (mpnn-#1-\name)
                    to[bend right=\MPNNMessageBend] (mpnn-#1-v);
                \draw[mpnn red message] (mpnn-#1-v)
                    to[bend right=\MPNNMessageBend] (mpnn-#1-\name);
            }
            % Both directions run outside the right edge to clear the spokes.
            \draw[mpnn blue message] (mpnn-#1-b.south east)
                .. controls +(\MPNNNeighbourInnerOffset,-.2)
                    and +(\MPNNNeighbourInnerOffset,.2) ..
                (mpnn-#1-d.north east);
            \draw[mpnn blue message] (mpnn-#1-d.east)
                .. controls +(\MPNNNeighbourOuterOffset,.3)
                    and +(\MPNNNeighbourOuterOffset,-.3) ..
                (mpnn-#1-b.east);
        \fi
    \end{scope}%
}

\begin{tikzpicture}[
    x=1cm, y=1cm, font=\small,
    line cap=round, line join=round,
    mpnn node/.style={
        circle, line width=.5pt, inner sep=0pt, outer sep=1pt,
        minimum size=\MPNNNodeSize
    },
    mpnn focus/.style={
        mpnn node, minimum size=\MPNNFocusSize,
        draw=figureRose, fill=figureRoseFill, text=figureRoseText
    },
    mpnn neighbour/.style={
        mpnn node, draw=figureBlue, fill=figureBlueFill
    },
    mpnn edge/.style={draw=black!65,line width=.45pt},
    mpnn arrow/.style={mpnn edge,draw=black,->,dashed,>=stealth},
    mpnn message/.style={
        mpnn arrow, dash pattern=on 1pt off 1pt, shorten >=1pt
    },
    mpnn blue message/.style={mpnn message,draw=figureBlue},
    mpnn red message/.style={mpnn message,draw=figureRose}
]
    \foreach \stage in {0,1,2} {\MPNNSnapshot{\stage}}

    % One shared message legend below the gap between the first two graphs.
    \node[below=2pt,text=black] (mpnn-messages)
        at ({.5*\MPNNRoundSeparation},\MPNNOutputHeight) {messages};
    \draw[mpnn blue message]
        ([xshift=-\MPNNLegendHalfWidth,yshift=2pt]mpnn-messages.north)
        .. controls +(.2,\MPNNLegendCurveHeight)
            and +(-.2,\MPNNLegendCurveHeight) ..
        ([xshift=\MPNNLegendHalfWidth,yshift=2pt]mpnn-messages.north);
    \draw[mpnn red message]
        ([xshift=\MPNNLegendHalfWidth,yshift=-2pt]mpnn-messages.south)
        .. controls +(-.2,-\MPNNLegendCurveHeight)
            and +(.2,-\MPNNLegendCurveHeight) ..
        ([xshift=-\MPNNLegendHalfWidth,yshift=-2pt]mpnn-messages.south);

    % Two updates between the three states.
    \foreach \stage in {0,1} {
        \pgfmathtruncatemacro{\LayerIndex}{\stage+1}
        \draw[mpnn arrow]
            ({\stage*\MPNNRoundSeparation+\MPNNHalfWidth+\MPNNUpdateStartGap},0) --
            ({(\stage+1)*\MPNNRoundSeparation-\MPNNHalfWidth-.33},0)
            node[midway,above=3pt] {$N_{\LayerIndex}$};
    }

    % The final classification violates P: v has four blue neighbours.
    \draw[mpnn arrow] (mpnn-2-v.south) --
        ({2*\MPNNRoundSeparation},\MPNNOutputHeight)
        node[below=2pt,text=black] {$M$ accepts};
\end{tikzpicture}
\endgroup
    \setlength{\abovecaptionskip}{2pt}
    \caption{MPNN $M$ with two layers ($N_1$ and $N_2$) that accepts node $v$.  
    Thus, $M$ is not safe regarding $P$. After each round the states of all nodes are updated (not visualised).}
    \label{sec:introduction;fig:verification}
\end{wrapfigure}

Given their prominence, recent years have seen substantial research on reasoning
about MPNNs, such as verifying their behaviour or characterising their expressive capabilities. 
In \emph{formal
verification}, we ask whether an MPNN satisfies a precisely specified property
$P$ on all admissible input graphs. For example, $P$ could require that ``a red
node is never accepted if it has more than three blue neighbours'', where
acceptance refers to binary classification.
Figure~\ref{sec:introduction;fig:verification} illustrates the processing of an MPNN $M$ with 
local message passing only, which violates the property $P$.

Such verification questions give rise to decision problems involving MPNNs -- 
a sound and complete verification procedure that always terminates decides the
corresponding problem. The complexity of this problem therefore constrains the
worst-case efficiency of such procedures. A representative problem is the so-called
\emph{satisfiability (or non-emptiness) problem of MPNNs}. 
It is defined as given a binary classifying MPNN $M$, decide whether $M$ 
accepts any pair $(G,v)$ of a graph
and node. Its simple form connects it directly to typical safety questions 
such as ``Is there any unwanted input that is accepted?'' or, conversely, 
``Can we ensure that no input of this kind is accepted?''. 
Typically, lower and upper complexity bounds for satisfiability carry over to such related problems 
or at least give indications. Thus, substantial research has investigated
satisfiability problems for MPNNs in recent years, both
directly
\cite{SalzerL23,Salzer25,SST25,BenediktLT26,ChernobrovkinSS26,LuTB26, benedikt2026decidingtwovariablelogicglobal}
and by relating their expressive capabilities to established formalisms with known complexity results
\cite{BarceloKM0RS20,NunnSST24,Grohe24,GrauFW26,SchonherrL26,SBGcT26,CucalaGMK23}.

The established results depend on the MPNN variant considered: whether aggregation
is local, global, or both, whether it uses sum, mean, or max to aggregate, and which
activation functions are used in the learnable FNN parts. On undirected graphs with
node labels from a finite set, satisfiability with exact rational arithmetic is
undecidable when local and global sum aggregation can be combined within a layer,
even with only truncated ReLU activations in the learnable FNN \cite{BenediktLT26}.
Restricting to local aggregation while keeping the remaining assumptions makes
satisfiability \textbf{PSPACE}-complete \cite{NunnSST24,BenediktLT26}.
Allowing both forms of aggregation, but in separate layers, still preserves
decidability with truncated ReLU activations
\cite{benedikt2026decidingtwovariablelogicglobal}.
Quantized MPNNs instead use finite-precision arithmetic, such as fixed- or
floating-point arithmetic, for both inputs and intermediate computations.
On directed graphs, satisfiability is \textbf{PSPACE}-complete with
local aggregation only \cite{SST25} and \NEXP{}-complete when global readout is allowed
\cite{ChernobrovkinSS26}.
Monotonicity provides another route to decidability: on graphs with Boolean node
features, monotonic max-sum MPNNs with ReLU activations and threshold classification
can be translated into positive Datalog with inequalities, yielding decidable
satisfiability \cite{CucalaGMK23}.
Without the monotonicity restriction, satisfiability is undecidable even for
local-sum MPNNs with ReLU activations on undirected coloured graphs
\cite{BenediktLT26}. 

Bounding the degree of valid input graphs, so considering the 
\emph{degree-bounded satisfiability problem (\dbsat)}, is another way to restore decidability 
even when arbitrary rational node labels are allowed \cite{SalzerL23}. However, with 
the degree bound given in unary, a gap was left between the known \textbf{PSPACE}-hardness 
and the \NEXP{} upper bound \cite[Theorems~6.28 and~6.29]{Salzer25}.
We close this gap by showing that \dbsat{} is in fact \NEXP-hard and, thus, with the
upper bound of \cite[Theorem~6.28]{Salzer25} it is fully specified as being
\NEXP-complete.
We remark that hardness already holds for the fixed degree bound $d=3$ and with
truncated ReLU activations. Compared with the \textbf{PSPACE} bounds for coloured
graphs and quantized MPNNs above, this highlights the role of rational input labels
whose precision remains unrestricted even when activations are bounded.

Our main result is established in Section~\ref{sec:mainresult}, through
Theorem~\ref{th:main} and formally stated in the following Corollary~\ref{cor:main}. Section~\ref{sec:overview}
outlines the proof and explains how the rational encodings, polyhedral conditions
and neural network constructions developed in Sections~\ref{sec:tm}--\ref{sec:fnn}
fit together.

\WFclear

\subsection*{AI Declaration}
The proof idea underlying this paper was proposed by GPT-6 Astra. The author
wrote the paper based on this idea, with further assistance from the same model
in checking and refining mathematical arguments, picking a fitting selection of related
work, revising the structure and wording, and writing TikZ code used for the figures.

\section{Proof Overview}
\label{sec:overview}
Our approach for proving \NEXP-hardness of \dbsat{} is standard: for each $\mathcal{P} \in \NEXP$, 
we establish a polynomial-time (many-one) reduction from $\mathcal{P}$ to 
\dbsat{}. For explanatory purposes, fix one such problem $\mathcal{P}$ in the following. 

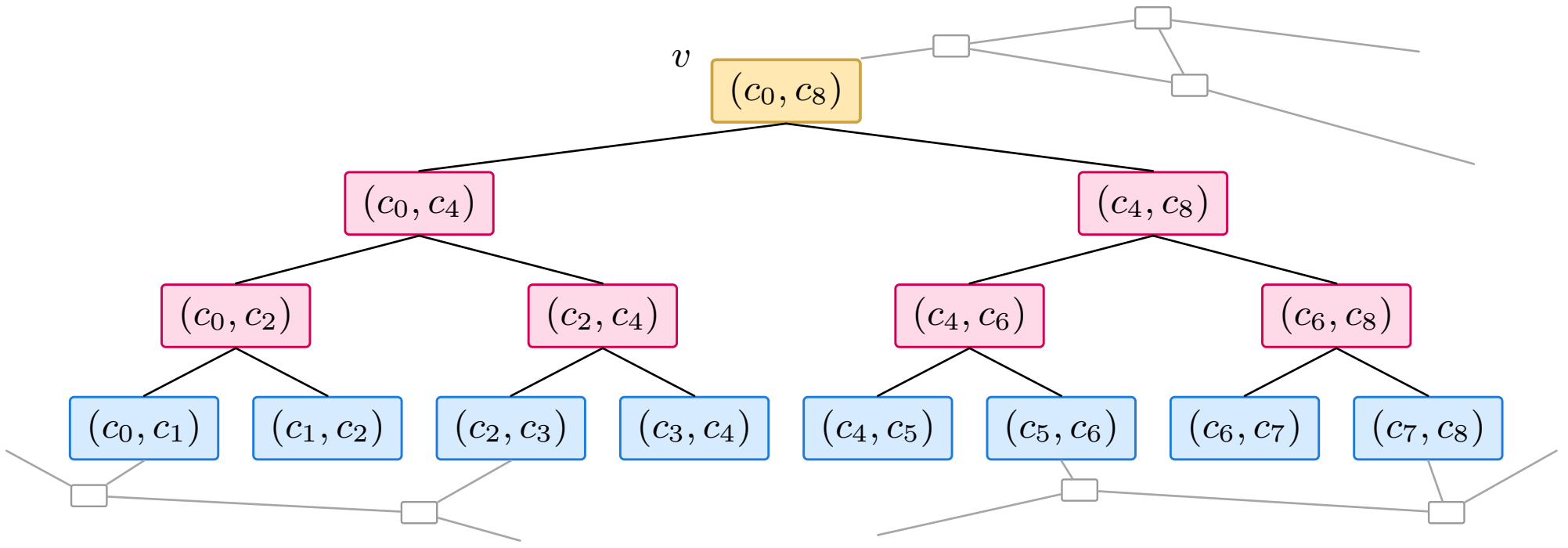
\begin{figure}
    \centering
    \begingroup
% Layout parameters: horizontal distance between leaves and vertical
% distance between successive levels. Styles are local to this picture.
\newcommand{\WitnessXSep}{1.55cm}
\newcommand{\WitnessYSep}{.95cm}

\begin{tikzpicture}[
    x=\WitnessXSep,
    y=\WitnessYSep,
    line cap=round,
    line join=round,
    witness node/.style={
        draw=black,
        fill=white,
        line width=.55pt,
        rounded corners=1pt,
        minimum width=1.13cm,
        minimum height=.47cm,
        inner xsep=4pt,
        inner ysep=3pt,
        font=\small
    },
    witness root/.style={witness node,figure focus border,fill=figureFocusFill},
    witness inner/.style={
        witness node,draw=figureRose,fill=figureRoseFill,text=figureRoseText
    },
    witness leaf/.style={witness node,draw=figureBlue,fill=figureBlueFill},
    witness edge/.style={draw=black,line width=.5pt},
    witness context node/.style={
        draw=black!40,fill=white,line width=.45pt,rounded corners=.5pt,
        minimum width=.3cm,minimum height=.18cm,inner sep=0pt
    },
    witness context edge/.style={draw=black!35,line width=.5pt},
    witness note/.style={font=\footnotesize,text=black!65}
]
    % The complete run and its two halves.
    \node[witness root] (v08) at ( 0, 0) {$(c_0,c_8)$};
    \node[font=\small,anchor=east] at ([xshift=-1pt]v08.north west) {$v$};
    \node[witness inner] (v04) at (-2,-1) {$(c_0,c_4)$};
    \node[witness inner] (v48) at ( 2,-1) {$(c_4,c_8)$};

    % Each half splits again at its intermediate configuration.
    \node[witness inner] (v02) at (-3,-2) {$(c_0,c_2)$};
    \node[witness inner] (v24) at (-1,-2) {$(c_2,c_4)$};
    \node[witness inner] (v46) at ( 1,-2) {$(c_4,c_6)$};
    \node[witness inner] (v68) at ( 3,-2) {$(c_6,c_8)$};

    % The leaves encode the eight successive single steps.
    \foreach \i in {0,...,7} {
        \pgfmathtruncatemacro{\j}{\i+1}
        \node[witness leaf] (leaf\i) at ({\i-3.5},-3)
            {$(c_{\i},c_{\j})$};
    }

    % Undirected graph edges; left/right position gives the segment order.
    \foreach \parent/\child in {
        v08/v04, v08/v48,
        v04/v02, v04/v24, v48/v46, v48/v68,
        v02/leaf0, v02/leaf1, v24/leaf2, v24/leaf3,
        v46/leaf4, v46/leaf5, v68/leaf6, v68/leaf7} {
        \draw[witness edge] (\parent.south) -- (\child.north);
    }

    % Surrounding graph, with small boxes at junctions. The root has one extra
    % neighbour; inner witness nodes already have degree 3. Omitted labels
    % near v must still satisfy the local checks, though they are not part
    % of the displayed decomposition. Every shown junction has degree <= 3.
    \node[witness context node] (context-top-a) at (.9,.4) {};
    \node[witness context node] (context-top-b) at (2,.65) {};
    \node[witness context node] (context-top-c) at (2.2,.05) {};
    \coordinate (context-top-d) at (3.45,.35);
    \coordinate (context-top-e) at (3.75,-.65);

    \node[witness context node] (context-left-a) at (-3.8,-3.6) {};
    \node[witness context node] (context-left-b) at (-2,-3.75) {};
    \coordinate (context-left-c) at (-4.25,-3.2);
    \coordinate (context-left-d) at (-1.45,-4);
    \node[witness context node] (context-right-a) at (1.6,-3.55) {};
    \node[witness context node] (context-right-b) at (3.6,-3.75) {};
    \coordinate (context-right-c) at (.5,-3.95);
    \coordinate (context-right-d) at (4.2,-3.2);

    \foreach \a/\b in {
        v08.north east/context-top-a,
        context-top-a/context-top-b, context-top-a/context-top-c,
        context-top-b/context-top-c, context-top-b/context-top-d,
        context-top-c/context-top-e,
        leaf0.south/context-left-a, leaf2.south/context-left-b,
        context-left-a/context-left-b, context-left-a/context-left-c,
        context-left-b/context-left-d,
        leaf5.south/context-right-a, leaf7.south/context-right-b,
        context-right-a/context-right-b, context-right-a/context-right-c,
        context-right-b/context-right-d} {
        \draw[witness context edge] (\a) -- (\b);
    }

    % \node[witness note,anchor=south,yshift=4pt] at (v08.north)
    %     {$c_0=c_{\mathrm{init}},\qquad c_8=c_{\mathrm{fin}}$};

    % Reading the leaves from left to right recovers the complete run.
    % \draw[
    %     draw=black!55,
    %     line width=.4pt,
    %     decorate,
    %     decoration={brace,mirror,amplitude=3pt}
    % ]
    %     ([yshift=-5pt]leaf0.south west) --
    %     ([yshift=-5pt]leaf7.south east)
    %     node[midway,below=8pt,font=\small,text=black] {
    %         $c_0\vdash_T c_1\vdash_T c_2\vdash_T c_3\vdash_T c_4
    %             \vdash_T c_5\vdash_T c_6\vdash_T c_7\vdash_T c_8$
    %     };
\end{tikzpicture}
\endgroup
    \caption{Graph $G$ with a highlighted tree subgraph $B$ rooted at $v$. The tree
        witnesses an accepting run $c_0,\dotsc,c_8$ of $T$ on $w$, with $c_0=c_w$
        and $c_8$ accepting.
        Each annotation $(c_i,c_j)$ represents the rational encodings of the
        endpoints of a run segment, stored in a node's label.}
    \label{sec:overview;fig:encoded_run}
\end{figure}

Given that $\mathcal{P} \in \NEXP$, there is a nondeterministic Turing machine (NTM)
$T$ that decides $\mathcal{P}$ in time at most $2^{p(n)}$ for some polynomial $p$.
Thus, for each $w \in \mathcal{P}$ there is an accepting run of $T$ on $w$ of at most exponential length and there is no such run if 
$w \notin \mathcal{P}$. To establish the reduction, we construct, for every $w$, a local-sum MPNN
$M_w$ in time polynomial in $|w|$ that accepts a graph-node pair $(G,v)$ of degree at most
three exactly when the pair witnesses an accepting run of $T$ on $w$ as shown in the example of $(G,v)$ 
in Figure~\ref{sec:overview;fig:encoded_run}.

The graph $G$ in Figure~\ref{sec:overview;fig:encoded_run} is of degree 3 and
contains a tree $B$ as a subgraph rooted at $v$. The label of each node of $B$
contains, among other auxiliary information, a rational encoding of a pair of
configurations $(c_i,c_j)$, representing a segment of the accepting run of $T$ on
$w$. The root $v$ contains the pair of start and accepting end configurations $(c_0,c_8)$,
the leaves contain pairs of immediate successor configurations $(c_i,c_{i+1})$,
and each inner node contains the start configuration of its left child and the
end configuration of its right child. The end configuration of the left child
must also agree with the start configuration of the right child. 
However, for arbitrary input graphs, $M_w$ cannot assume that these label coordinates
actually encode configurations. We therefore write $(\bs{x},\bs{y})$ for the pair of
vectors in the following.

The key point is that the local conditions for leaf, inner and root nodes together
ensure that the stored pairs describe a binary decomposition of an accepting run.
This decomposition need not be a tree subgraph of $G$: different branches may share
nodes, whose occurrences we treat as separate copies when unfolding the decomposition
into a tree.
We start with standard but well-behaved definitions of NTM configurations and
corresponding rational encodings $\enc$, established in Section~\ref{sec:tm}, which
then allow us to define the relation $\Delta$ between vectors in
Section~\ref{sec:accept} with the following central property: if $(\bs{x},\bs{y}) \in \Delta$ and
$\bs{x} = \enc(c)$, meaning it is a valid encoding of a configuration $c$, then $\bs{y} = \enc(c')$, where $c'$ is a valid
successor configuration of $c$, or $c'=c$ and $c$ is accepting. The latter case
allows padding of accepting runs. This is reflected by Lemma~\ref{sec:tm;lem:bounded_acceptance}.

Correctness mainly follows from an inductive invariant, established from
leaves to root: at each node, the tuple $(\bs{x},\bs{y})$ represents a valid, possibly
padded run segment whenever all checks in its subtree succeed and $\bs{x}$ encodes
a valid configuration. At a leaf, $M_w$ must check that $(\bs{x},\bs{y}) \in \Delta$,
so the base case follows from the property of $\Delta$ described above. At an inner
node, the local checks require the two children to store pairs $(\bs{x},\bs{z})$
and $(\bs{z},\bs{y})$. Starting from a valid encoding $\bs{x}$, the induction
hypothesis for the left child gives a valid run segment ending at a valid encoding
$\bs{z}$. This lets us apply the induction hypothesis to the right child and
concatenate the two segments, establishing the invariant at their parent.
At the root, the first endpoint is fixed to $\enc(c_w)$, a valid encoding of the
initial configuration on $w$, and the last endpoint must have the accepting state.
Applying the invariant at the root therefore shows that the decomposition witnesses
an accepting run of $T$ on $w$.

By construction, $\Delta$ is a finite union of rational polyhedra
(see Equation~\ref{eq:delta}) and thus recognisable by FNNs with truncated ReLU
activations, as we show in Section~\ref{sec:fnn}. This means membership in $\Delta$
can be checked by the learnable part of $M_w$.
The check for inner nodes can also be performed in a single message passing round.
For this, each node's label specifies its level in the tree and whether it is a left
or right child. Its endpoint pair is stored in a block of
coordinates specific to this combination, together with a marker set to $1$, while
all other blocks must be zero. Summation then keeps the contributions of different
levels and child positions separate and, for valid labels, the markers count the
neighbours of each kind. The root must have level $0$, and each inner node must have
exactly one left and one right child at the next level. Their endpoint pairs can
thus be recovered from the sum and checked for matching endpoints. Nodes at level
$p(|w|)$ are leaves.
Since $M_w$ depends on $w$, it can easily check that the root stores
$\bs{x} = \enc(c_w)$. The condition that its final endpoint has the accepting state can
also be checked by testing membership in a finite union of rational polyhedra,
namely $F$ defined in Section~\ref{sec:accept}.

To collect the results of the local checks, $M_w$ stores an error in $[0,1]$ at
each node, zero exactly when its local conditions hold. In subsequent rounds,
each node adds its neighbours' errors to its own and truncates the sum at $1$.
Non-negative errors cannot cancel, and positive errors remain positive under
truncation.
Here, the binary decomposition of a run as a tree is important: after padding an
accepting run to length $2^{p(|w|)}$, recursively splitting it into halves yields a
tree of height $p(|w|)$ and degree at most three. Thus, although the run may be
exponentially long, $M_w$ can propagate the errors to the root $v$ in $p(|w|)$
message passing rounds. It accepts exactly when the accumulated error at $v$ is zero
and the root conditions hold.

Since $T$ is fixed, $\Delta$ and $F$ have constant-size polyhedral descriptions.
There are only polynomially many label blocks and linear conditions, and
$\enc(c_w)$ can be computed in polynomial time with polynomial bit length.
Thus, the checking FNN can be constructed in polynomial time, as can the remaining
polynomially many layers of $M_w$.

These ingredients are combined in Theorem~\ref{th:main}, which relates satisfiability
of $M_w$ on graphs of degree at most three to the polyhedral condition from
Lemma~\ref{sec:tm;lem:bounded_acceptance}. By this lemma, the condition is equivalent
to acceptance of $w$ by $T$. Thus, $w \mapsto (M_w,3)$ gives the required
polynomial-time reduction, yielding the hardness result in Corollary~\ref{cor:main}.

\section{Turing Machines and Rational Encodings of Configurations}
\label{sec:tm}
We work with a standard definition of a \emph{nondeterministic Turing machine (NTM)}
$T$ given as a tuple $T = (\Gamma, \Sigma, Q, \delta, q_0, \square, q_F)$, where
$\Gamma$ is the finite \emph{working alphabet}, $\Sigma \subset \Gamma$ is the
\emph{input alphabet}, and $Q$ is the finite set of \emph{states}. The relation
$\delta \subseteq (Q \times \Gamma) \times (Q \times \Gamma \times \{L,R\})$ is the
\emph{transition relation}, $q_0 \in Q$ is the \emph{start state},
$\square \in \Gamma \setminus \Sigma$ is the \emph{blank symbol}, and $q_F \in Q$
is the \emph{final state}.

\begin{figure}
    \centering
    \begingroup
% Layout: cell dimensions and distances from the tape to the annotations.
% All commands and styles are local to this picture.
\newcommand{\TMCellWidth}{1cm}
\newcommand{\TMCellHeight}{.46cm}
\newcommand{\TMHeadOffset}{.85cm}
\newcommand{\TMReadOffset}{.52cm}

\begin{tikzpicture}[
    x=\TMCellWidth,
    line cap=round,
    line join=round,
    tm cell/.style={
        draw=black, fill=none, text=black, line width=.45pt,
        minimum width=\TMCellWidth, minimum height=\TMCellHeight,
        inner sep=0pt, outer sep=0pt, font=\small
    },
    tm blank cell/.style={
        tm cell, draw=figureRose, fill=figureRoseFill, text=figureRoseText
    },
    tm written cell/.style={
        tm cell, draw=figureBlue, fill=figureBlueFill
    },
    tm state/.style={
        figure focus border, fill=none, rounded corners=2pt,
        inner xsep=6pt, inner ysep=3pt, font=\small
    },
    tm arrow/.style={->, figure focus border},
    tm label/.style={font=\small}
]
    % Tape contents, listed in their physical order from left to right.
    % Position 0 is reserved for the scanned symbol.
    % Blank tails fade outwards; entries give distance from the head / opacity.
    \foreach \distance/\fade in {7/.25,6/.6,5/1} {
        \foreach \side in {-1,1} {
            \node[tm blank cell, opacity=\fade]
                at ({\side*\distance},0) {$\square$};
        }
    }
    \foreach \position/\symbol in {
        -4/a_m, -2/a_2, -1/a_1, 1/b_1, 2/b_2, 4/b_n} {
        \node[tm written cell] at (\position,0) {$\symbol$};
    }
    \foreach \position in {-3,3} {
        \node[tm written cell] at (\position,0) {$\dotsb$};
    }
    \node[tm cell, figure focus border, fill=figureFocusFill] (scanned) at (0,0) {$a$};

    % The machine state and the head pointing to its scanned cell.
    \node[tm state] (state) at (0,\TMHeadOffset) {state $q$};
    \draw[tm arrow] (state.south) -- (scanned.north);

    % Reading order of the configuration words, from the head outwards.
    % The arrows span only the finite words, not their blank tails.
    % \draw[tm arrow] (-.6,-\TMReadOffset) -- (-4.4,-\TMReadOffset)
    %     node[midway,below=3pt,tm label] {$l=a_1a_2\dotsb a_m$};
    % \draw[tm arrow] (.6,-\TMReadOffset) -- (4.4,-\TMReadOffset)
    %     node[midway,below=3pt,tm label] {$r=b_1b_2\dotsb b_n$};
\end{tikzpicture}
\endgroup
    \caption{Visualisation of an NTM configuration $c=(l,(q,a),r)$ with $l= a_1 \dotsb
        a_m$ and $r = b_1 \dotsb b_n$.}
    \label{sec:tm;fig:configuration}
\end{figure}
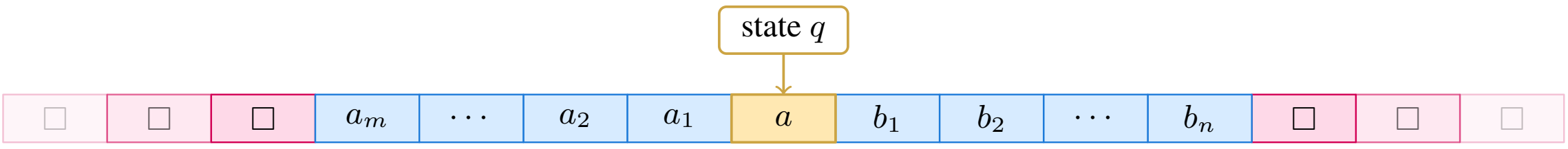

We call a triple $c=(l, (q,a), r) \in (\Gamma^*) \times (Q \times \Gamma) \times
\Gamma^*$ a \emph{configuration} of $T$ and interpret it as follows: $T$ is in state
$q$ and its head is on the symbol $a$. The words $l$ and $r$ list the tape symbols to
the left and right of the head, respectively, ordered from nearest to farthest.
Thus, if $l=a_1 \dotsb a_m$, that part of the tape reads $a_m \dotsb a_1$ from left
to right. We interpret $l$ and $r$ as representing
$l\square^\omega$ and $r\square^\omega$. This is visualised in
Figure~\ref{sec:tm;fig:configuration}.

Let $aw \in \Gamma^+$ and define $\mathrm{head}(aw) = a$, $\mathrm{head}(\epsilon)=\square$ and $\mathrm{tail}(aw)=w$,
$\mathrm{tail}(\epsilon)=\epsilon$. Assuming $T$ is in configuration $(l, (q,a), r)$,
and there is a transition $((q,a),(q',a',L)) \in \delta$, then a possible next
configuration is $(\mathrm{tail}(l), (q',\mathrm{head}(l)), a'r)$. Analogously, if we
have a transition $((q,a),(q',a',R)) \in \delta$, a possible next configuration is
$(a'l, (q',\mathrm{head}(r)), \mathrm{tail}(r))$. If for two configurations $c_i$ and
$c_j$
%there is a step of $T$ from $c_i$ to $c_j$, we write $c_i \vdash_T c_j$, if
there is a step given by transition $t$ we write $c_i \vdash_t c_j$, and if there is a
sequence of such steps, called a \emph{run}, we write $c_i \vdash^*_T c_j$. 
Given a run $c_0, \dotsc, c_n$, we call $n$ its 
\emph{length} and call the run \emph{accepting} if $c_n = (w, (q_F,a), w')$ for some $w, w'
\in \Gamma^*$, and $a \in \Gamma$. Accordingly, we call configuration $c_n = (w,
(q_F,a), w')$ an \emph{accepting configuration}.
For a nonempty input $w=av$ with $a\in\Sigma$, the initial configuration is
$c_w=(\epsilon,(q_0,a),v)$ and for the empty word $w=\epsilon$ it is
$c_\epsilon=(\epsilon,(q_0,\square),\epsilon)$.

\subsection*{Rational Encodings of Configurations}

We assume from now on w.l.o.g.\ that the working alphabet of $T$ is $\Gamma \subset \nats$ with
symbols  $1, \dotsc, k$. Let $b = k+2$ and $I = [\frac{1}{b-1},\frac{k}{b-1}]$. We
define $\mathrm{enc}(a_1 \dotsb a_m) = \sum_{i=1}^m a_i b^{-i} + \frac{s}{(b-1)b^m}$ for
all $a_1 \dotsb a_m \in \Gamma^*$, where $s$ is the number representing the blank
symbol\footnote{In fact, $\mathrm{enc}(a_1 \dotsb a_m)$ encodes the infinite word $a_1
\dotsb a_m \square^\omega$ implying that $\epsilon$ is encoded as $\square^\omega$.}.
Let $c = (l, (q,a), r)$ be a configuration of $T$. We denote by $\enc(c) \in [0,1]^3$
the three-dimensional vector 
\begin{displaymath}
\enc(c) = (\enc(l), \enc(q,a), \enc(r)), 
\end{displaymath}
where $\enc(q,a)$ is an
arbitrary unique rational encoding in $[0,1]$ for all pairs $(q,a) \in Q \times \Gamma$.
Furthermore, for all $a \in \Gamma$, we define
$\push_a,\pop_a \colon \reals \rightarrow \reals$ by $\push_a(x) = \frac{1}{b}(x+a)$
and $\pop_a(x) = bx-a$.

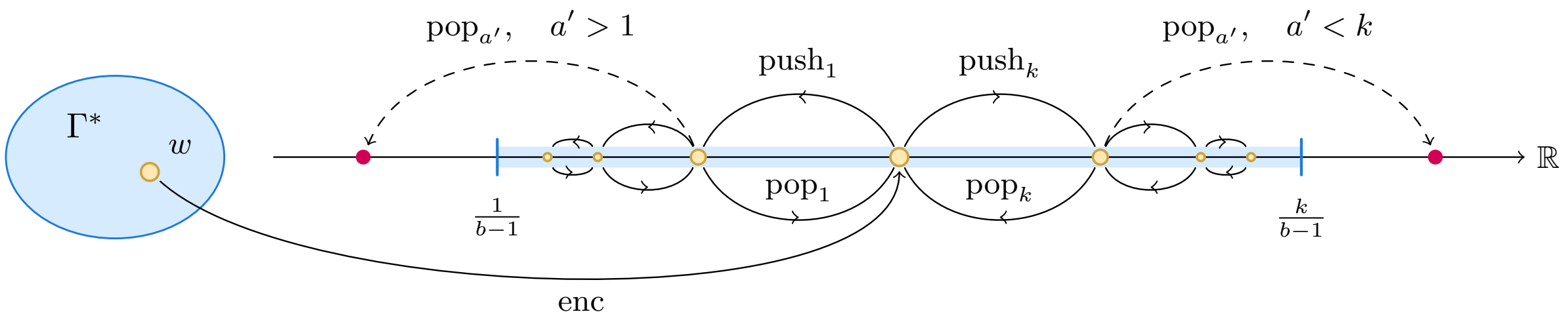
\begin{figure}[t]
    \centering
    \begingroup
% Schematic positions; the hop spacing is chosen for legibility, not to scale.
% The two limiting encodings are the endpoints of I.
\newcommand{\EncodingLower}{5}
\newcommand{\EncodingUpper}{13.1}
\newcommand{\EncodingWord}{9.05}
\newcommand{\EncodingLeftOutside}{3.65}
\newcommand{\EncodingRightOutside}{14.45}
\newcommand{\EncodingBoundaryHeight}{.18}
\newcommand{\EncodingHopCount}{3}
\newcommand{\EncodingHopGap}{3.5pt}
\newcommand{\EncodingHopArrowPosition}{.5}
\newcommand{\EncodingHopArrowScale}{1.05}

\begin{tikzpicture}[
    x=1cm, y=1cm, font=\small,
    line cap=round, line join=round,
    encoding arrow/.style={
        ->,draw=black,line width=.45pt,shorten >=1.5pt,shorten <=1.5pt
    },
    encoding hop/.style={
        draw=black,line width=.45pt,
        shorten >=\EncodingHopGap,shorten <=\EncodingHopGap,
        postaction={decorate},
        decoration={markings,mark=at position \EncodingHopArrowPosition with
            {\arrow[scale=\EncodingHopArrowScale]{>}}}
    },
    encoding escape/.style={encoding arrow,dashed},
    encoding point/.style={circle,fill=black,inner sep=1.5pt,outer sep=0pt},
    encoding small point/.style={encoding point,inner sep=.8pt},
    encoding focus point/.style={
        encoding point,inner sep=1.8pt
    },
    encoding inside/.style={figure focus border,fill=figureFocusFill},
    encoding outside/.style={fill=figureRose},
    encoding domain/.style={draw=figureBlue,fill=figureBlueFill,line width=.55pt},
    encoding interval/.style={draw=figureDomainFill,line width=6pt,line cap=butt},
    encoding boundary/.style={draw=figureBlue,line width=.8pt}
]
    % (1) Encodings lie in the highlighted interval I = [1/(b-1),k/(b-1)].
    \draw[encoding interval]
        (\EncodingLower,0) -- (\EncodingUpper,0);
    \draw[->,draw=black,line width=.45pt]
        (2.75,0) -- (15.35,0) node[right] {$\reals$};
    \foreach \position/\symbol in
        {\EncodingLower/{\frac{1}{b-1}},\EncodingUpper/{\frac{k}{b-1}}} {
        \draw[encoding boundary]
            (\position,\EncodingBoundaryHeight) -- (\position,-\EncodingBoundaryHeight)
            node[below=3pt,text=black] {$\symbol$};
    }

    % (2) Outward pushes approach the limits; the reverse hops are valid pops.
    % Mid-curve arrowheads leave the shared dots clear.
    % Only the first pair of arrows on either side is labelled.
    \foreach \side/\symbol in {-1/1,1/k} {
        \pgfmathsetmacro{\HopExtent}{\side<0 ?
            \EncodingWord-\EncodingLower : \EncodingUpper-\EncodingWord}
        \foreach \i in {1,...,\EncodingHopCount} {
            \pgfmathsetmacro{\PreviousX}
                {\EncodingWord+\side*\HopExtent*(1-pow(.5,\i-1))}
            \pgfmathsetmacro{\NextX}{\EncodingWord+\side*\HopExtent*(1-pow(.5,\i))}
            \pgfmathsetmacro{\HopWidth}{abs(\NextX-\PreviousX)}
            \pgfmathsetmacro{\HopHeight}{.4*\HopWidth}
            \pgfmathsetmacro{\HopControl}{\side*.2*\HopWidth}
            \draw[encoding hop] (\PreviousX,0)
                .. controls +(\HopControl,\HopHeight) and +(-\HopControl,\HopHeight) ..
                (\NextX,0);
            \draw[encoding hop] (\NextX,0)
                .. controls +(-\HopControl,-\HopHeight) and
                +(\HopControl,-\HopHeight) ..
                (\PreviousX,0);
            \ifnum\i=1
                \node[encoding point,encoding inside] (first\side) at (\NextX,0) {};
                \node[above=2pt] at ({(\PreviousX+\NextX)/2},{.75*\HopHeight})
                    {$\push_{\symbol}$};
                \node at ({(\PreviousX+\NextX)/2},-.34) {$\pop_{\symbol}$};
            \else
                \node[encoding small point,encoding inside] at (\NextX,0) {};
            \fi
        }
        \pgfmathsetmacro{\DotsX}
            {\EncodingWord+\side*\HopExtent*(1-pow(.5,\EncodingHopCount+1))}
        %\node[font=\scriptsize,inner sep=0pt] at (\DotsX,0) {$\cdots$};
    }

    % A central label and a light guide keep it clear of the first two hops.
    % \draw[draw=black!40,line width=.35pt]
    %     (\EncodingWord,.12) -- (\EncodingWord,1.02)
    %     node[above=2pt,text=black] {$\enc(w)$};
    \node[encoding focus point,encoding inside]
        (encoded) at (\EncodingWord,0) {};

    % The domain stays beside the axis; its map passes beneath the hop chains.
    \draw[encoding domain] (1.15,0)
        ellipse[x radius=1.1cm,y radius=.82cm];
    \node[font=\normalsize] at (.85,.32) {$\Gamma^*$};
    \node[encoding focus point,encoding inside,label=above right:{$w$}]
        (word) at (1.5,-.15) {};
    \draw[encoding arrow] (word)
        .. controls +(1.5,-1.35) and +(0,-1.65) ..
        node[midway,below=2pt] {$\enc$} (encoded);

    % (3) Use the first point on each side, whose head is 1 or k respectively.
    \node[encoding point,encoding outside] (outside-left)
        at (\EncodingLeftOutside,0) {};
    \node[encoding point,encoding outside] (outside-right)
        at (\EncodingRightOutside,0) {};
    \draw[encoding escape] (first-1)
        .. controls +(-.5,1.25) and +(.5,1.25) ..
        node[midway,above=2pt,text=black] {$\pop_{a'},\quad a'>1$} (outside-left);
    \draw[encoding escape] (first1)
        .. controls +(.5,1.25) and +(-.5,1.25) ..
        node[midway,above=2pt,text=black] {$\pop_{a'},\quad a'<k$} (outside-right);
\end{tikzpicture}
\endgroup
    \caption{Schematic illustration of the properties (1), (2), and (3) of configuration
        encodings provided by Lemma~\ref{sec:tm;lem:enc_operations}.}
    \label{sec:tm;fig:enc_operations}
\end{figure}

To define the relation $\Delta$ already outlined in Section~\ref{sec:overview}, we need
arithmetic tape updates that preserve valid encodings. The following lemma
provides us with these operations and a test for whether the symbol being popped
is correct: starting from a valid tape encoding, $\push_a$ gives the encoding of
the extended tape, while $\pop_a$ stays in $I$ exactly when $a$ is the symbol
nearest to the head. These properties are illustrated in Figure~\ref{sec:tm;fig:enc_operations}.

\begin{lemma}
    \label{sec:tm;lem:enc_operations}
    Let $w \in \Gamma^*$ and $a \in \Gamma$. We have that
    \begin{enumerate}
        \item[(1)] $\enc(w) \in I = [\frac{1}{b-1},\frac{k}{b-1}]$,
        \item[(2)] $\push_a(\enc(w)) = \enc(aw)$ and $\pop_{\mathrm{head}(w)}(\enc(w)) =
            \enc(\mathrm{tail}(w))$, and
        \item[(3)] $\pop_a(\enc(w)) \in I$ iff $a = \mathrm{head}(w)$.
    \end{enumerate}
\end{lemma}
\begin{proof}
    Since every symbol of $\Gamma$ is between $1$ and $k$, we have
    $\frac{1}{b-1} = \sum_{i=1}^{\infty} b^{-i} \leq \enc(v) \leq \sum_{i=1}^{\infty} k
    b^{-i} = \frac{k}{b-1}$ for every $v \in \Gamma^*$. This proves (1). 
    By definition of $\enc$, we obtain $\enc(av)=\frac{a+\enc(v)}{b}$.
    For the empty word, we have
    $\enc(\epsilon)=\frac{s}{b-1}=\frac{s+\enc(\epsilon)}{b}$.
    Setting $h=\mathrm{head}(w)$ and $z=\enc(\mathrm{tail}(w))$ gives
    $\enc(w)=\frac{h+z}{b}$, including when $w=\epsilon$.
    The push identity in (2) follows from
    $\push_a(\enc(w))=\frac{a+\enc(w)}{b}=\enc(aw)$. Moreover,
    $\pop_a(\enc(w))=b\cdot\enc(w)-a=h+z-a$.
    For $a=h$, this equals $z$, proving the pop identity in (2), and belongs to $I$
    by (1). If $a\neq h$, then $h-a$ is a nonzero integer, so $h+z-a \notin I$,
    since $z\in I$ and $I$ has width $\frac{k-1}{b-1}<1$. This proves (3).
\end{proof}

% Place the illustration at the top of the following page.
\suppressfloats[t]

\section{A Polyhedral Characterisation of Bounded NTM Acceptance}
\label{sec:accept}
% We use the rational encodings of configurations $\enc(c)$ and the operations
% $\push_a$ and $\pop_a$ defined in Section~\ref{sec:tm} to describe valid steps of
% NTM $T$ as polyhedra. Composing the resulting polyhedra then allows us to express
% acceptance within a given number of steps.

A \emph{polyhedron} $P \subseteq \reals^n$ is defined by $P = \{\bs{x} \mid A \bs{x}
\leq \bs{b}\}$ where $A \in \reals^{m\times n}$ and $\bs{b} \in \reals^m$, meaning it is a
subset of $\reals^n$ defined by a linear inequality system. We call $P$ \emph{rational}
if $A$ and $\bs{b}$ can be chosen with rational entries. Similarly, we can see the
condition $A \bs{x} \leq \bs{b}$ as a conjunction of $m$ linear inequalities. Taking
this view, we freely use linear equality conditions to specify polyhedra as well. These 
equalities are simply represented by two linear inequalities.

We use the encoding $\enc$ and the operations $\push_a$, $\pop_a$ to define a relation
$\Delta$, given by a finite union of rational polyhedra, with the following property:
starting from a valid configuration encoding, $\Delta$ describes a step of $T$ or a
repetition of an accepting configuration. Its iterated composition then allows us to
characterise bounded acceptance.

For each transition $t = ((q_1,a_1),(q_2,a',d)) \in \delta$ and $a_2 \in \Gamma$, we
define $R_{t,a_2}$ for $d=R$ and $L_{t,a_2}$ for $d=L$. To allow repetitions of
accepting configurations, we also define $P_a$ for each $a \in \Gamma$:
\begin{align*}
    R_{t,a_2} &= \{((x_1,h_1,y_1), (x_2,h_2,y_2)) \mid x_i,y_i \in I, h_i =
        \enc(q_i,a_i), x_2 = \push_{a'}(x_1), y_2 = \pop_{a_2}(y_1)\}, \\
    L_{t,a_2} &= \{((x_1,h_1,y_1), (x_2,h_2,y_2)) \mid x_i,y_i \in I, h_i =
        \enc(q_i,a_i), y_2 = \push_{a'}(y_1), x_2 = \pop_{a_2}(x_1)\}, \\
    P_a &= \{((x,h,y),(x,h,y)) \mid x,y \in I, h = \enc(q_F,a)\}.
\end{align*}
Here, $i \in \{1,2\}$. The conditions in $R_{t,a_2}$ and $L_{t,a_2}$ describe the update
of a triple $(x_1,h_1,y_1) \in [0,1]^3$ under a right or left transition $t$,
respectively, assuming that the symbol under the head is $a_2$ afterwards. Note that
these triples are not necessarily valid encoded configurations of $T$.

Each $R_{t,a_2}$, $L_{t,a_2}$ and $P_a$ is a rational polyhedron: the interval $I$ has
rational endpoints, the encodings of state-symbol pairs are fixed rational numbers,
and $\push$ and $\pop$ are affine maps with rational coefficients. There are
$|\delta| \cdot |\Gamma|$ transition polyhedra and $|\Gamma|$ polyhedra $P_a$. We define
\begin{align}
    \label{eq:delta}
    \Delta = \bigcup_{a \in \Gamma} \bigl(P_a \cup \bigcup_{t\in \delta_r}R_{t,a} \cup
        \bigcup_{t \in \delta_l} L_{t,a}\bigr),
\end{align}
where $\delta_r$ and $\delta_l$ denote the subsets of right and left moving transitions
of $\delta$. Given previous arguments, $\Delta$ is a finite union of $|\Gamma| \cdot
(|\delta|+1)$ rational polyhedra. 

For $i \in \nats$, we denote by $\Delta^i$ the \emph{$i$-fold
composition} of $\Delta$, defined by $(\bs{x},\bs{y}) \in \Delta^i$ iff there are
$\bs{x}_0,\dotsc,\bs{x}_i \in \reals^3$ with $\bs{x}_0=\bs{x}$, $\bs{x}_i=\bs{y}$, and
$(\bs{x}_{j-1},\bs{x}_j) \in \Delta$ for all $1 \leq j \leq i$. In particular,
$\Delta^0$ is the identity relation.
We also define
$F = \bigcup_{a \in \Gamma} \{(x,\enc(q_F,a),y) \mid x,y \in I\}$.

\begin{lemma}
    \label{sec:tm;lem:bounded_acceptance}
    Let $T$ be an NTM, $w$ an input word, and $n \in \nats$. There is an accepting run
    of $T$ on $w$ of length at most $n$ if and only if $(\enc(c_w),\bs{x}) \in \Delta^n$
    for some $\bs{x} \in F$.
\end{lemma}
\begin{proof}
    Let $c$ be the final configuration of an accepting run $r$ of length $i \leq n$
    starting at $c_w$. Lemma~\ref{sec:tm;lem:enc_operations} provides that $\push$ and
    $\pop$ as they are used in the definition of $R_{t,a}$ and $L_{t,a}$ capture the
    semantics of a step of $T$, and the construction of $\Delta$ then ensures that each
    step of $r$ corresponds to a pair in $\Delta$. Thus, for its $i$-fold composition we
    have $(\enc(c_w),\enc(c)) \in \Delta^i$ with $\enc(c) \in F$. If $i < n$, the sets
    $P_a \subseteq \Delta$ allow us to pad $r$ by repeating $\enc(c)$.

    For the other direction, let $\bs{x} \in F$ with $(\enc(c_w),\bs{x}) \in \Delta^n$.
    We first show that $\Delta$ preserves valid configuration encodings: if $c$ is a
    configuration and $(\enc(c),\bs{z}) \in \Delta$, then either $\bs{z}=\enc(c')$ for
    some successor configuration $c'$ of $c$, or $c$ is accepting and $\bs{z}=\enc(c)$.
    To see this, write $c=(l,(q,a),r)$. If $(\enc(c),\bs{z}) \in R_{t,a_2}$ for some $t
    \in \delta$ and $a_2 \in \Gamma$, the unique encoding of state-symbol pairs ensures
    that $t$ is applicable to $c$. Moreover, $\pop_{a_2}(\enc(r)) \in I$ forces
    $a_2=\mathrm{head}(r)$ by Lemma~\ref{sec:tm;lem:enc_operations}.3, and both tape
    coordinates of the target are then the correct encodings by
    Lemma~\ref{sec:tm;lem:enc_operations}.2. The argument for $L_{t,a_2}$ is symmetric,
    while a pair in $P_a$ simply repeats an accepting configuration. Starting at the
    valid encoding $\enc(c_w)$, it follows by induction that the witnessing chain
    consists of valid steps and repetitions of accepting configurations. Its endpoint
    $\bs{x}$ is therefore a valid configuration encoding, and membership in $F$ means
    that this configuration is accepting. Removing the padding steps gives an accepting
    run of length at most $n$.
\end{proof}

We say that $T$ is \emph{time-bounded by $f(n)$}, for $f \colon \nats \to \nats$, if for
all $w \in \Sigma^*$ all possible runs of $T$ starting with configuration $c_w$ are of
length at most $f(|w|)$. If $T$ is time-bounded by $f(n)$,
Lemma~\ref{sec:tm;lem:bounded_acceptance} with $n=f(|w|)$ gives
\begin{align*}
    T \text{ accepts } w && \text{if and only if} &&
    (\enc(c_w),\bs{x}) \in \Delta^{f(|w|)} \text{ for some } \bs{x} \in F.
\end{align*}

\section{Recognising Finite Unions of Polyhedra with Truncated ReLU-FNNs}
\label{sec:fnn}
% In this section, we argue that each finite union of polyhedra is efficiently
% recognisable by a Feedforward Neural Network (FNN) with truncated ReLU activations.
% Given
% Lemma~\ref{sec:tm;lem:bounded_acceptance}, we can thus use an FNN to check acceptance of
% a word $w$ by an NTM $T$.

An FNN $N$ with truncated ReLU activations is a composition of layers built from
truncated ReLU nodes. A \emph{truncated ReLU node} with input dimension $m\in\nats$
computes a function $v:\reals^m\to\reals$ of the form
$v(\bs{x})=\trelu(b+\sum_{i=1}^m w_i x_i)$, where
$\trelu(y)=\min\{1,\max\{0,y\}\}$, $\bs{x}=(x_1,\ldots,x_m)$,
$w_i\in\mathbb{Q}$ is the weight on input $i$, and $b\in\mathbb{Q}$ is the bias.
A \emph{layer} $\ell:\reals^m\to\reals^n$ consists of $n$ such nodes, each taking
the same $m$-dimensional input. A \emph{feedforward neural network (FNN)} with $k$
layers is then defined by
$N=\ell_k\circ\cdots\circ\ell_1:\reals^{m_1}\to\reals^{m_{k+1}}$, where
$\ell_j:\reals^{m_j}\to\reals^{m_{j+1}}$ for $j=1,\ldots,k$.

\begin{lemma}
    \label{lem:check_polyhedral_cond}
    For each finite union of rational polyhedra $S \subseteq \reals^m$ with $S =
    \bigcup_{i=1}^n P_i$ there is an FNN $N_S$ such that for all $\bs{x} \in \reals^m$
    we have $N_S(\bs{x}) \in [0,1]$ and $N_S(\bs{x}) = 0$ iff $\bs{x} \in S$.
\end{lemma}
\begin{proof}
    For the empty union, take the constant network $N_S=1$. Otherwise, let $P_i =
    \{\bs{x} \in \reals^m \mid A_i\bs{x} \leq \bs{b}_i\}$, where $A_i \in
    \rats^{m_i\times m}$ and $\bs{b}_i \in \rats^{m_i}$ for $1 \leq i \leq n$, and the
    inequality is understood componentwise. For $\bs{x} \in \reals^m$, we define
    \begin{displaymath}
        \rho_{P_i}(\bs{x}) = \trelu\left(
            \sum_{j=1}^{m_i} \trelu\bigl((A_i\bs{x}-\bs{b}_i)_j\bigr)\right),
        \qquad
        \rho_S(\bs{x}) = \min_{1 \leq i \leq n} \rho_{P_i}(\bs{x}),
    \end{displaymath}
    where $(\bs{y})_j$ denotes the $j$-th component of a vector $\bs{y}$.
    Each $\rho_{P_i}$ takes values in $[0,1]$ and equals $0$ iff all inequalities
    defining $P_i$ hold. Thus, $\rho_S(\bs{x}) \in [0,1]$ and
    $\rho_S(\bs{x})=0$ iff $\bs{x} \in S$.

    Each $\rho_{P_i}$ is realised by two truncated ReLU layers: the first computes
    the summands and the second their sum truncated at $1$. The minimum of two
    values $x,y\in[0,1]$ can be realised by a truncated ReLU FNN as well, via
    $\min(x,y) = \trelu(x - \trelu(x-y))$. Values in $[0,1]$ can be passed unchanged
    through additional layers since $\trelu(x)=x$ on this interval. Computing all
    $\rho_{P_i}$ in parallel and successively applying this construction yields
    the required FNN $N_S=\rho_S$.
\end{proof}

Furthermore, the proof of Lemma~\ref{lem:check_polyhedral_cond} yields an FNN
recognising $S$ whose size is polynomial in the number of polyhedra $P_i$ and the total
representation size of their defining rational parameters $A_i,\bs{b}_i$. Here and
throughout, we define the size of an FNN as the sum of the representation sizes of its
weights and biases.

Truncated ReLU can also be realised by a constant-size ReLU-FNN, since
$\trelu(y)=\relu(\relu(y)-\relu(y-1))$, where $\relu(y)=\max\{0,y\}$.
Replacing each truncated ReLU node by this construction yields an ordinary
ReLU-FNN of polynomial size.

\section{\NEXP-hardness of \dbsat{} for Local-Sum MPNNs}
\label{sec:mainresult}
A \emph{local-sum message passing neural network (MPNN)} $M$ is a tuple
$(N_1,\ldots,N_k)$ of FNNs with truncated ReLU activations, where
$N_i:\reals^{2d_{i-1}}\to\reals^{d_i}$ for $i=1,\ldots,k$ and $d_0,\ldots,d_k\in\nats$.
Given an undirected graph $G=(V,E,L)$ with node set $V$, edge set $E$, and rational
node labels $L:V\to\rats^{d_0}$, $M$ assigns each node $v\in V$ the initial state
$\bs{x}_v^{(0)}=L(v)$ and updates its state in layer $i$ by
\[
    \bs{x}_v^{(i)} = N_i\left(\bs{x}_v^{(i-1)},
    \sum_{\{u,v\}\in E}\bs{x}_u^{(i-1)}\right).
\]
Thus, each update uses only the node's own state and the sum of its neighbours' states.
Empty sums denote zero vectors of the appropriate dimension. For a \emph{pointed
graph} $(G,v)$, consisting of a graph $G$ and a designated node $v\in V$, we write
$M(G,v)=\bs{x}_v^{(k)}$ for the final state of $v$. We say that $M$ \emph{accepts}
$(G,v)$ iff $M(G,v)_{d_k} = 1$.

We define the \emph{degree of a graph $G$} as $\max_{u \in V} |\{w \in V \mid \{u,w\} \in E\}|$.
The \emph{degree-bounded satisfiability problem} of local-sum MPNNs, denoted by $\dbsat$,
is to decide, given a local-sum MPNN $M$ and a degree bound $d\in \nats$ encoded in unary,
whether there is a pointed graph $(G,v)$ of degree at most $d$ such that $M$ accepts
$(G,v)$.

Our goal is to show that $\dbsat$ is \NEXP-hard, meaning that every decision problem in
\NEXP{} has a polynomial-time many-one reduction to $\dbsat$.
Let $\text{NTIME}(f(n))$ be the class of decision problems decided by an NTM in time
$\mathcal{O}(f(n))$, and let $\NEXP = \bigcup_{k \geq 1} \text{NTIME}(2^{n^k})$.
Thus, \NEXP{} is the class of decision problems solvable in nondeterministic
exponential time.
To establish hardness, fix an arbitrary problem $\mathcal{P} \in \NEXP$ and an NTM $T$
deciding $\mathcal{P}$ within $2^{p(n)}$ steps for some polynomial
$p \colon \nats \to \nats$. 
% The core insight is that, using the rational polyhedral
% description developed in the previous sections, we can efficiently construct a
% local-sum MPNN $M_w$ that checks whether a pointed graph witnesses an accepting run
% of $T$ on $w$.
\begin{theorem}
    \label{th:main}
    For all $w \in \Sigma^*$ there is a local-sum MPNN $M_{w}$ with truncated ReLU
    activations such that the following are equivalent:
    \begin{enumerate}
        \item There is a pointed graph $(G,v)$ of degree at most 3 that is accepted by
            $M_{w}$.
        \item There is $b \in F$ with $(\enc(c_w), b) \in \Delta^{2^{p(|w|)}}$.
    \end{enumerate}
    Furthermore, $M_{w}$ can be constructed in time polynomial
    in $|w|$.
\end{theorem}
\begin{proof}[Proof]
    Set $h=p(|w|)$, and let $(G,v)$ be an arbitrary pointed graph of degree at most
    $3$ given as input to $M_w$. We construct $M_w$ such that it accepts $(G,v)$ iff
    the pointed graph $(G,v)$ witnesses $(\enc(c_w),b) \in \Delta^{2^h}$ for some $b \in F$, as
    illustrated in Figure~\ref{sec:overview;fig:encoded_run}. In particular, the label
    of $v$ contains $(\enc(c_w),b)$ with $b \in F$. The labels of its descendants
    recursively split each pair into two pairs with matching endpoints. After $h$
    levels, the leaf pairs must belong to $\Delta$. Here, we adopt tree terminology, 
    although the input graph need not be a tree. 
    The endpoint checks are illustrated in Figure~\ref{sec:mainresult;fig:local_checks}.

    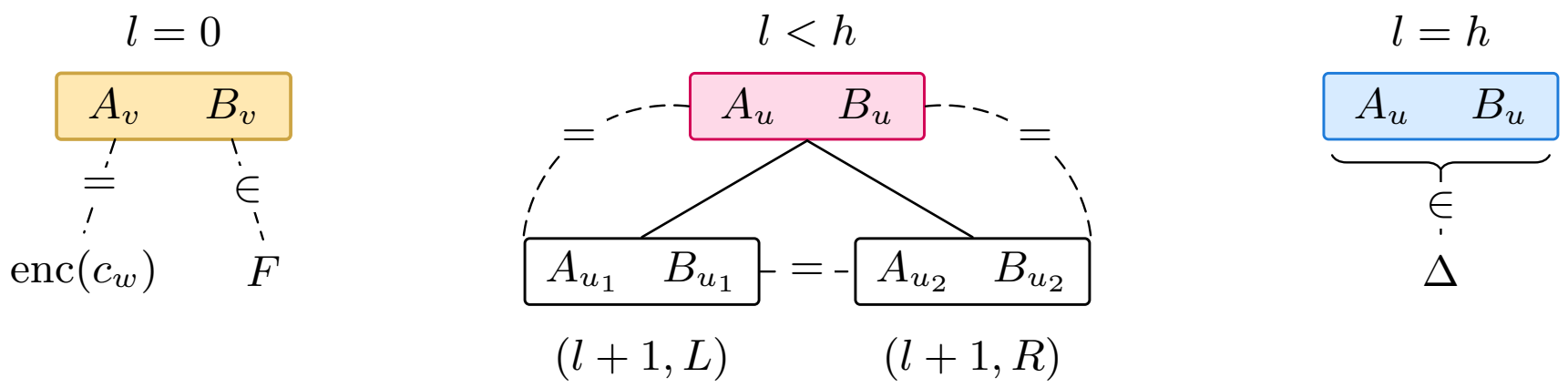
\begin{figure}[t]
        \centering
        \begingroup
% Panel spacing and node geometry; all definitions are local to this figure.
\newcommand{\ChecksPanelSep}{4.6cm}
\newcommand{\ChecksChildSep}{1.2cm}
\newcommand{\ChecksLevelSep}{1.2cm}
\newcommand{\ChecksNodeWidth}{1.7cm}
\newcommand{\ChecksNodeHeight}{.48cm}
\newcommand{\ChecksBowAngle}{45}

% A node contains two endpoints in one rectangle. Named anchors attach checks
% below each endpoint, while solid graph edges attach to the enclosing box.
\newcommand{\ChecksPair}[4]{%
    \node[#1] (#2) at (0,0) {};
    \node[check value] at (-.25*\ChecksNodeWidth,0) {$#3$};
    \node[check value] at (.25*\ChecksNodeWidth,0) {$#4$};
    \coordinate (#2-A) at (-.25*\ChecksNodeWidth,-.5*\ChecksNodeHeight);
    \coordinate (#2-B) at (.25*\ChecksNodeWidth,-.5*\ChecksNodeHeight);
}

\begin{tikzpicture}[
    x=1cm, y=1cm, line cap=round, line join=round,
    check node/.style={
        draw=black, fill=white, text=black, line width=.55pt,
        rounded corners=1pt, minimum width=\ChecksNodeWidth,
        minimum height=\ChecksNodeHeight, inner xsep=5pt, inner ysep=3pt,
        font=\small
    },
    check root/.style={check node,figure focus border,fill=figureFocusFill},
    check inner/.style={
        check node,draw=figureRose,fill=figureRoseFill,text=figureRoseText
    },
    check leaf/.style={check node,draw=figureBlue,fill=figureBlueFill},
    check edge/.style={draw=black,line width=.5pt},
    check relation/.style={draw=black,densely dashed,line width=.45pt},
    check symbol/.style={font=\small,text=black,fill=white,inner sep=2pt},
    check value/.style={font=\small,inner sep=0pt},
    check label/.style={font=\small,text=black},
    check type/.style={font=\footnotesize,text=black}
]
    % Root: the first endpoint is initial and the second is accepting.
    \begin{scope}
        \node[check type] at (0,.55) {$l=0$};
        \ChecksPair{check root}{root}{A_v}{B_v}
        \node[check label] (initial) at (-.65cm,-\ChecksLevelSep)
            {$\enc(c_w)$};
        \node[check label] (accepting) at (.65cm,-\ChecksLevelSep) {$F$};
        \draw[check relation] (root-A) -- node[check symbol] {$=$}
            (initial.north);
        \draw[check relation] (root-B) -- node[check symbol] {$\in$}
            (accepting.north);
    \end{scope}

    % Inner node: one child of each required type, with matching endpoints.
    % Children are neutral: at the next level they may be inner nodes or leaves.
    \begin{scope}[xshift=\ChecksPanelSep]
        \node[check type] at (0,.55) {$l<h$};
        \begin{scope}[text=figureRoseText]
            \ChecksPair{check inner}{inner}{A_u}{B_u}
        \end{scope}
        \begin{scope}[xshift=-\ChecksChildSep,yshift=-\ChecksLevelSep]
            \ChecksPair{check node}{left}{A_{u_1}}{B_{u_1}}
        \end{scope}
        \begin{scope}[xshift=\ChecksChildSep,yshift=-\ChecksLevelSep]
            \ChecksPair{check node}{right}{A_{u_2}}{B_{u_2}}
        \end{scope}
        \draw[check edge] (inner.south) -- (left.north);
        \draw[check edge] (inner.south) -- (right.north);
        % The bows meet the top corners without crossing into the child boxes.
        \draw[check relation] (inner.west) to[bend right=\ChecksBowAngle]
            node[check symbol,midway] {$=$} (left.north west);
        \draw[check relation] (inner.east) to[bend left=\ChecksBowAngle]
            node[check symbol,midway] {$=$} (right.north east);
        \draw[check relation] (left.east) -- node[check symbol] {$=$}
            (right.west);
        \node[check type,anchor=north] at ([yshift=-3pt]left.south) {$(l+1,L)$};
        \node[check type,anchor=north] at ([yshift=-3pt]right.south) {$(l+1,R)$};
    \end{scope}

    % Leaf: the endpoint pair is a transition or an accepting padding step.
    \begin{scope}[xshift={2*\ChecksPanelSep}]
        \node[check type] at (0,.55) {$l=h$};
        \ChecksPair{check leaf}{leaf}{A_u}{B_u}
        \draw[decorate,decoration={brace,mirror,amplitude=3pt},line width=.45pt]
            ([xshift=2pt,yshift=-3pt]leaf.south west) --
            ([xshift=-2pt,yshift=-3pt]leaf.south east);
        \node[check label] (transition) at (0,-\ChecksLevelSep) {$\Delta$};
        \draw[check relation] (0,-.47cm) -- node[check symbol] {$\in$}
            (transition.north);
    \end{scope}
\end{tikzpicture}
\endgroup
        \caption{Endpoint checks for valid node types used in the proof of Theorem~\ref{th:main}. The root also satisfies the
            inner or leaf check.}
        \label{sec:mainresult;fig:local_checks}
    \end{figure}

    To do this, we construct $M_w$ such that it maintains two error flags in $[0,1]$
    at each node. In the first layer, $(\bs{x}_u^{(1)})_1$ is $0$ iff the
    local encoding conditions around $u$ hold, while $(\bs{x}_u^{(1)})_2$ is $0$
    iff $u$ satisfies the special conditions required of the root. These flags are
    computed at every node, but among the second flags, only the one at $v$ matters for acceptance.
    In the next $h$ layers, each node adds its neighbours' first flags to its
    own, truncates the sum at $1$, and keeps its second flag unchanged. Truncation
    preserves positivity, so all local errors within distance $h$ reach $v$ by
    layer $h+1$. Finally, the last layer $N_{h+2}$ computes
    \begin{displaymath}
        M_w(G,v) = \trelu\bigl(1-(\bs{x}_v^{(h+1)})_1
            -(\bs{x}_v^{(h+1)})_2\bigr).
    \end{displaymath}
    Hence, $M_w$ accepts $(G,v)$ iff both flags at $v$ are $0$.

    In detail, we construct $M_w$ as follows. We use types
    $t=(l,x) \in \{0,1,\dotsc,h\} \times \{L,R\}$, where $l$ denotes the level in
    the decomposition and $x$ indicates a left or right child. At level $0$, either
    value of $x$ is allowed.
    The initial labelling of a valid input graph $G = (V,E,L)$ must be of the form $L
    \colon V \to \rats^d$ with $d = 2(h+1)\cdot 7$. This is simply enforced by
    constructing $N_1$ with input dimension $2d$. We interpret each label as a series of
    blocks $(z_t,A_t,B_t)$ with $z_t \in \reals$ and $A_t,B_t \in \reals^3$, one for
    each type $t$. We require that each node $u \in V$ has exactly one type $t$,
    meaning that
    \begin{align*}
        z_t = 1, A_t,B_t \in [0;1]^3 && \text{and} && z_{r}=0, A_{r}=B_{r}=\bs{0}
            \text{ for all } r \neq t.
    \end{align*}
    For a node $u$ of type $t$, we write $z_u = z_t$, $A_u = A_t$, and $B_u = B_t$
    for its active label block. We use primes to denote the corresponding blocks
    of the aggregated label $\sum_{\{u,u'\}\in E} L(u')$. Provided all neighbours
    have valid types, $z'_t$ counts the neighbours of type $t$, since inactive
    blocks are zero. If $z'_t=1$, then $A'_t$ and $B'_t$ are exactly the endpoints
    stored by that neighbour.

    If node $u$ has type $t=(l,x)$ with $l<h$, we additionally require
    \begin{align*}
        z'_{(l+1,L)} = 1, z'_{(l+1,R)} = 1,
            && A_u = A'_{(l+1,L)}, B_u = B'_{(l+1,R)},
            && \text{and} && B'_{(l+1,L)} = A'_{(l+1,R)}.
    \end{align*}
    If $l=h$, we instead require $(A_u,B_u) \in \Delta$. The conditions for an inner
    node $u$ identify exactly one child of type $(l+1,L)$ and one of type $(l+1,R)$,
    provided its neighbours have valid types.
    The $L$ child $u_1$ must start at $A_u$, the $R$ child $u_2$ must end at $B_u$,
    and their intermediate endpoints must agree, meaning $B_{u_1}=A_{u_2}$.
    Thus, if the two children encode partial runs, these concatenate to a partial run
    from $A_u$ to $B_u$.

    For each type, the type conditions together with the respective child or leaf
    conditions define a finite union of rational polyhedra in the input space of $N_1$.
    Taking the union over all types, Lemma~\ref{lem:check_polyhedral_cond} allows us
    to realise these checks via $N_1$ such that $(\bs{x}^{(1)}_u)_1$ lies in $[0,1]$
    and equals $0$ iff all local conditions hold. Additionally, we use $N_1$ to
    enforce that $(\bs{x}^{(1)}_u)_2$ lies in $[0,1]$ and equals $0$ iff $u$ has type
    $(0,x)$ for some $x \in \{L,R\}$, $A_u=\enc(c_w)$, and $B_u \in F$. These root
    conditions also define a finite union of rational polyhedra, so
    Lemma~\ref{lem:check_polyhedral_cond} applies again. Realising both checks in
    parallel is done by $N_1$.

    Assume that $M_w$ accepts $(G,v)$. Then all local conditions hold within
    distance $h$ of $v$, and the root conditions hold at $v$. Starting at $v$, we
    follow the children prescribed by the local conditions. A node reached at
    level $l<h$ is at distance at most $l$ from $v$, so all its neighbours are
    within distance $h$ and have valid types. Thus, the aggregated coordinates
    identify exactly one child of each required type. The following induction does
    not require nodes reached along different branches to be distinct. By induction
    from level $h$ upwards, each reached node $u$ at level $l$ satisfies
    $(A_u,B_u) \in \Delta^{2^{h-l}}$: this holds directly at the leaves, and at an
    inner node the two child pairs compose because their intermediate endpoints
    agree. At the root, we therefore obtain $(\enc(c_w),B_v) \in \Delta^{2^h}$ with
    $B_v \in F$.

    Conversely, suppose that $(\enc(c_w),b) \in \Delta^{2^h}$ for some $b \in F$.
    By definition, there is a chain of $2^h$ steps in $\Delta$ from $\enc(c_w)$ to
    $b$, with all entries in $[0,1]^3$.
    We construct a full binary tree of height $h$, whose root stores the endpoints
    of this chain. Recursively, the two children store the endpoints of its first
    and second halves. We assign types according to the level and child position,
    choosing $(0,L)$ for the root, and set all inactive blocks to zero. The leaves
    then store pairs in $\Delta$. All local conditions hold, and the root conditions
    hold at $v$. All labels are rational, since $\enc(c_w)$ is rational and each
    step in $\Delta$ uses rational affine updates. The tree has degree at most $3$
    and $M_w$ accepts this pointed graph.

    Finally, constructing $M_w$ takes time polynomial in $|w|$. Since $T$ is fixed,
    $\Delta$ and $F$ each consist of a constant number of rational polyhedra.
    There are $2(h+1)$ types, each giving a constant number of polyhedra with
    $\mathcal{O}(h+1)$ linear conditions for the first flag. The root conditions
    also give a constant number of polyhedra. All their coefficients are fixed
    except for the entries of $\enc(c_w)$, which can be computed in polynomial
    time and have polynomial bit length in $|w|$. Thus, the construction from
    Lemma~\ref{lem:check_polyhedral_cond} gives an FNN $N_1$ of polynomial size.
    The remaining $h+1$ layers have constant size, and $h=p(|w|)$ is polynomial
    in $|w|$.
\end{proof}
Using Lemma~\ref{sec:tm;lem:bounded_acceptance}, we have that
Theorem~\ref{th:main} constructs for each $w \in \Sigma^*$ a local-sum MPNN $M_w$ in
polynomial time that accepts some pointed graph of degree at most $3$ if and only
if $w \in \mathcal{P}$. Now, mapping $w$ to $(M_w,3)$, which can be done in polynomial
time in $|w|$, gives the required reduction from $\mathcal{P} \in \NEXP$ to $\dbsat$.
\begin{corollary}
    \label{cor:main}
    For local-sum MPNNs with truncated ReLU activations, $\dbsat$ is
    \NEXP-hard even when the degree bound is fixed to $3$.
\end{corollary}
The same complexity bounds hold for ordinary ReLU activations, since the
construction can be translated into a ReLU-MPNN of polynomial size using the
simulation in Section~\ref{sec:fnn}.

\section{Conclusion}
\label{sec:conclusion}

We showed that $\dbsat$ for local-sum MPNNs with truncated ReLU activations over
graphs with rational node labels is \NEXP-hard, even when the degree bound is fixed
to $3$.
Together with the known upper bound \cite[Theorem~6.28]{Salzer25}, this establishes
\NEXP-completeness when the degree bound is given in unary.

For coloured graphs with labels from $\{0,1\}^m$, or a fixed finite rational label
set, the same degree-bounded problem is in \textbf{PSPACE}. A witness tree can be
guessed and checked on the fly, keeping only a branch of polynomial length together
with the relevant node states and partial neighbour sums. These values have
polynomial bit length, so the procedure uses nondeterministic polynomial space,
which equals \textbf{PSPACE}.

For unrestricted rational labels, this argument lacks a polynomial bound on the
bit length of witness labels. Our construction exploits this freedom to encode
configurations with arbitrarily long tape contents. The MPNN checks transitions
between these encodings locally and combines the results by sum aggregation to
verify an accepting run of at most exponential length. This combination yields
\NEXP-hardness even with input label coordinates in $[0,1]$ and bounded activations.

\section*{Acknowledgements}
I thank Martin Lange for discussions when we first stumbled upon this problem in \cite{SalzerL23}.

\bibliographystyle{plainnat}
\bibliography{bib/references}

% \appendix
% \input{appendices/proofs}

\end{document}